\documentclass[usenatbib]{mn2e}
\usepackage{txfonts}
\usepackage{graphicx}

\title[Thin accretion discs around millisecond X-ray pulsars]{Thin accretion discs around millisecond X-ray pulsars}
\author[Solomon Belay Tessema and Ulf Torkelsson]{Solomon Belay Tessema$^{1,2}$\thanks{E-mail:
newtonsolbel@yahoo.com; torkel@physics.gu.se} and Ulf Torkelsson
$^{2}$\\
$^{1}$Department of Physics,Addis Ababa University, P.O.Box 1176, Addis Ababa, Ethiopia\\
$^{2}$Department of Physics,  University of Gothenburg, SE 412 96 Gothenburg, Sweden}
\begin{document}

\date{Accepted . Received ; in original form }

\pagerange{\pageref{firstpage}--\pageref{lastpage}} \pubyear{2009}

\maketitle

\label{firstpage}

 \begin{abstract}
Millisecond x-ray pulsars have weak magnetic dipole moments of
$\sim 10^{16}$\,T\,m$^3$ compared to ordinary X-ray pulsars with dipole moments
of $10^{20}$\,T\,m$^3$.  For this reason a surrounding accretion disc can
extend closer to the neutron star, and thus reach a higher temperature,
at which the opacity is dominated by electron scattering and radiation
pressure is strong.
We compute the self-similar structure of such a geometrically thin axisymmetric
accretion disc with an internal dynamo.
Such models produce significantly stronger torques on the neutron star than
models without dynamos, and can explain the strong spin variations in some
millisecond X-ray pulsars.
 \end{abstract}

\begin{keywords}
 accretion, accretion discs -- magnetic fields -- MHD --stars: magnetic fields
-- stars: neutron -- X-rays:binaries -- pulsars: general
 \end{keywords}

\section{Introduction}
The first accretion-powered
millisecond pulsar, SAX J1808.4-3658, was discovered a
decade ago
\citep{wk,chm}.
Since then several other millisecond X-ray pulsars with spin periods between 1.7
and 5.4\,ms have been discovered
(e.g. XTE J0929-314, XTE J1751-305, XTE 1807-294 and XTE
J1814-338; see \citealt{wij} for a review).  These neutron
stars have considerably weaker magnetic dipole moments,
$\sim 10^{16}$\,T\,m$^3$, than the conventional X-ray pulsars.
The accretion disc then extends much
closer to the neutron star, where it reaches such a high temperature that the
opacity is dominated by electron scattering and radiation pressure can
grow stronger than the gas pressure.
On the other hand the accretion rates in these systems can be so low that the
Alfv\'en radius might be located outside of the co-rotation radius of the
neutron star.  Conventionally it has been thought that such systems will
eject the mass that is transferred from the companion star
through the propeller mechanism \citep{ill75}, though it has been pointed out
by \citet{st} and \citet{rfs} that the velocity excess right outside the
co-rotation radius is insufficient to eject all the matter.
\citet{st} found that when the Alfv\'en radius is only slightly larger
than the co-rotation radius then the disc will cycle between one state in
which the matter is accumulated outside
the co-rotation radius and another state in which it is accreted onto the
star.
Later \citet{rfs} constructed a steady accretion disc model, in which the entire
disc is located outside of
the co-rotation radius.  The magnetic field of the neutron star is penetrating
this disc and is mediating an exchange of angular momentum between the
accretion disc and the neutron star as in the model by \citet{gb}.
Since the entire accretion
disc is rotating more slowly than the neutron star in this case,
the disc brakes the
rotation of the neutron star.

The models of \citet{gb} and \citet{rfs} have ignored the possibility of an
internal dynamo in the accretion disc, and have considered only the
magnetic field that is generated in the disc by the mismatch in the rotation
of the neutron star and the disc.  However the magnetic field due to a
dynamo can strengthen the coupling between the accretion disc and the
neutron star, and we \citep{tt} have shown that the angular momentum
exchange between the accretion disc and the neutron star can increase
by more than an
order of magnitude compared to the standard model by
\citet{gb}.  Furthermore the dynamo is insensitive to the rotation of
the neutron star, and can therefore in principle provide a spin-up torque
even at a low accretion rate.

A complementary approach to study the magnetic interaction between
an accretion disc and a star is represented by
the numerical simulations by \citet{lon}.
In these simulations it is possible to study more complicated magnetic field
geometries than
are accessible to semi-analytic studies (see for instance \citealt{lon}),
but the resolution is so low that it is not possible to study the
appearance of small-scale turbulence and the resulting turbulent transport
and dynamo action in the simulation.  Thus these simulations do not really
address the viability of a dynamo in the disc.  However there are other things
one can learn from these simulations.  They show how the matter is streaming
via funnel flows from the accretion disc onto the magnetic poles of the neutron
star.  Furthermore they show that angular momentum is also lost
from the accretion disc through the jets that are accelerated away from the
disc.  \citet{rom2009} make a distinction between a conical jet that transport
most of the mass, and a less dense axial jet inside the conical jet.  The latter
component can dominate the energetics if the star is in the propeller regime.
It is also noteworthy that these simulations confirm that some matter might
still be accreted when the star is in the propeller regime.

The purpose of this paper is to extend our previous study (Paper I)
of an accretion disc with an internal dynamo around a magnetised star
to the discs around millisecond pulsars. The difference is that due to the
weak magnetic field the disc extends to much smaller radii, at which it becomes
so hot that a significant part of the opacity is due to electron scattering,
and radiation pressure might dominate over the gas pressure.
Therefore in the spirit of
\citet{bs,bs1} we divide the disc into an outer region with
gas pressure and free-free opacity, a middle region with gas pressure and
electron scattering opacity, and an inner region with radiation pressure and
electron scattering.

In Sect. 2 we derive a single ordinary differential equation, which is generally
applicable for the radial
structure of the accretion disc independently of the equation of state and the
form of the
opacity. We then introduce the equations of state and the opacity prescriptions
for the different regions of the
disc in Sect. 3
and present the numerical solutions in Sect. 4. The properties of these solutions
and their application to the millisecond X-ray pulsars are discussed in Sect.
5, and we finally summarise our results in Sect. 6.

\section{Governing equations}

\subsection{The magnetohydrodynamic angular momentum balance}

Our approach is an extension of the standard model for a thin accretion disc
(\citealt{bs}), and it follows closely the method that we introduced in
Paper I, though we will now consider a general opacity law and
equation of state.  For this reason we only summarise the steps leading up
to Eq. (21) of Paper I, after which we deviate from the treatment in Paper I.

The surface density of the accretion disc is defined as
\begin{equation}\label{1}
  \Sigma = \int_{-\infty}^\infty \rho \mbox{d}z = 2\rho H,
\end{equation}
where $\rho$ is the density, and $H$  the half thickness of the disc.  We
can then write the accretion rate as
\begin{equation}\label{2}
  \dot M = - 2\pi R \Sigma v_R,
\end{equation}
which is independent of the radial coordinate $R$, and where $v_R$ is the
radial velocity.

A thin accretion disc is supported by the centrifugal force in the radial
direction, so we get that
  \begin{equation}\label{3}
    v_{\mathrm{\phi}}^{2} - \frac{GM}{R} = 0,
   \end{equation}
where $v_\phi$ is the azimuthal velocity, $G$ the gravitational
constant and $M$ is the mass of the accretor.   This shows that the disc
rotates in a Keplerian fashion.

The disc is in hydrostatic equilibrium in the vertical direction, which
gives us that the pressure at the midplane
    of the disc is
   \begin{equation}\label{4}
        P = \frac{1}{2}H\Sigma\frac{GM}{R^{3}} = \rho\frac{GMH^{2}}{R^{3}}
   \label{hydro-eq}
   \end{equation}
   but the hydrostatic equilibrium can also be expressed as
   \begin{equation}\label{5}
        \frac{H}{R} = \frac{c_{\mathrm s}}{v_{\mathrm {kepl}}}
   \end{equation}
where
   \begin{equation}\label{6}
        c_{\mathrm {s}} = \left(\frac{P}{\rho}\right)^{1/2}
   \end{equation}
   is the isothermal speed of sound.  This shows that the Keplerian velocity is
highly supersonic in a thin accretion disc.

The conservation of angular momentum leads to the following height-integrated
equation (Paper I)
\begin{equation}\label{ang_mom}
    \Sigma \left(v_{R}\frac{\mbox{d} l}{\mbox{d} R}\right) =
     \left[\frac{B_{z}B_{\phi}}{\mu_{\mathrm{0}}}\right]_{-H}^{H}R  +
\frac{1}{R}\frac{\mbox{d}}{\mbox{d} R}\left[R^{3}\nu\Sigma
\frac{\mbox{d}}{\mbox{d} R}
    \left(\frac{l}{R^{2}}\right)\right],
\end{equation}
where
the specific angular momentum $l = Rv_{\phi }\propto R^{1/2}$ and $\nu$ is
the viscosity.
The magnetic term describes the exchange of angular momentum between the disc
and the star via the magnetosphere.
We assume that the vertical magnetic field is due to the dipolar field of
the neutron star, so that its value in the equatorial plane is
\begin{equation}\label{8}
    B_{z} = -\frac{\mu}{R^{3}}
  \label{dipole}
 \end{equation}
where $\mu$ is the magnetic dipole moment of the star.
There are two sources for $B_\phi$.  The shear between the Keplerian disc and
the magnetosphere produces a field whose value in the upper half of the disc is
\begin{equation}\label{9}
 B_{\mathrm{\phi,shear}} = -\gamma B_{z}\left(1 -
\frac{\Omega_{\mathrm{s}}}{\Omega_{\mathrm{k}}}\right),
\label{shear}
 \end{equation}
where  $\Omega_{\rm k}= v_{\phi}/R,$\, $\Omega_{\mathrm {s}}$ is the angular
velocity of the star,  and
$\gamma$ is a dimensionless parameter of the order of a few
\citep{gb}.

The magnetohydrodynamical turbulence in the accretion disc is also
generating a magnetic field, $B_{\phi, {\rm dyn}}$,
through a dynamo action (see for instance \citealt{bh2}).  In order
to estimate the strength of this magnetic field we start by making the
assumptions that the turbulence is not affected by $B_{\phi,{\rm shear}}$,
which is justified because in general $|B_{\phi, {\rm turb}}| >
|B_{\phi,{\rm shear}}|$ \citep{tt}, and that the accretion is driven in its
entirety by the small-scale turbulent magnetic (Maxwell) stress.
The latter assumption ignores the fact that the Reynolds stress is
responsible for roughly 20\% of the radial transport of angular momentum
\citep{brandenburg2}, but this results in only a minor quantitative change.
It is then the $R\phi$-component of the turbulent Maxwell stress tensor,
$\langle B_{R,{\rm turb}} B_{\mathrm{\phi,turb}}\rangle/\mu_0$, that
transports the angular momentum, and we couple this to the pressure in the
disc through the \citet{bs} $\alpha_{\rm ss}$-prescription
\begin{equation}\label{10}
 f_{R\phi} = \frac{-\left\langle B_{R,{\rm turb}}
B_{\mathrm{\phi,turb}}\right\rangle}{\mu_{\mathrm{0}}} =
\alpha_{\mathrm{ss}}P(r).
\end{equation}
Note that a negative Maxwell stress means that the angular momentum is
transported outwards and that numerical simulations suggest that
$\alpha_{\rm ss} \sim 10^{-2}$ \citep{haw95,brandenburg2}.
\citet{tor98} suggested that\footnote{This is a more precise definition
of $\gamma_{\rm dyn}$ than $B_\phi = \gamma_{\rm dyn} B_r$ that was
used by \citet{tor98}.}
 \begin{equation}\label{11}
\gamma_{\mathrm {dyn}} =
\frac{\left\langle B_{\phi,{\rm turb}}^2\right\rangle}
{-\left\langle B_{R,{\rm turb}}B_{\phi,{\rm turb}}\right\rangle},
 \end{equation}
where $\gamma_{\mathrm {dyn}}\sim 10$ based on the numerical simulations by
\citet{brandenburg2}.  We now assume that due to an inverse cascade this
turbulent filed contains
a large-scale field, $B_{\phi,{\rm dyn}}$ \citep{fri75}, and we express
the ratio of the large-scale field to the rms-value of the turbulent field as
\begin{equation}
  \epsilon = \frac{B_{\phi,{\rm dyn}}}
{\left\langle B_{\phi,{\rm turb}}^2\right\rangle^{1/2}}.
\end{equation}
Thus we can express the large-scale magnetic field, that couples to the
magnetic field of the neutron star, as
\begin{equation}\label{12}
       B_{\mathrm\phi,{\rm dyn}} = \epsilon\left(\alpha_{\rm ss}\mu_{0}\gamma_{dyn}P(r)\right)^{1/2},
\label{b-dyn}
\end{equation}
where
$-1 \leq \epsilon \leq 1$,
and a negative value describes a magnetic field which is pointing in the
negative
$\phi$-direction at the upper disc surface.  We will in general be conservative
and assume that $|\epsilon| \leq 0.1$, but the simulations by
\citet{brandenburg2} suggest that $|\epsilon|$ can be significantly larger.

We can now re-express Eq. (\ref{ang_mom}) as
 \begin{eqnarray}\label{ang-mom1}
    \Sigma \left(v_{R}\frac{\mbox{d} l}{\mbox{d} R}\right) =
     2\epsilon\frac{B_{z}}{\mu_{\mathrm{0}}}\left(\alpha_{\mathrm{ss}}\mu_{\mathrm{0}}\gamma_{\mathrm{dyn}}P(r)\right)^{1/2}R  - 2\gamma\frac{B^{2}_{z}}{\mu_{\mathrm{0}}}\frac{\left(\Omega_{\mathrm{k}}- \Omega_{\mathrm{s}}\right)}{\Omega_{\mathrm{k}}}R\nonumber\\
      + \frac{1}{R}\frac{\mbox{d}}{\mbox{d} R}\left[R^{3}\nu\Sigma
\frac{\mbox{d}}{\mbox{d} R}
    \left(\frac{l}{R^{2}}\right)\right].
\end{eqnarray}
which can be formulated as an ordinary differential equation in $\nu\Sigma$
(Paper I).

\subsection{Heating and radiative transport}

For a slow inflow of matter through an optically thick disc the local viscous
dissipation  $\mathbf{v\cdot f_{\mathrm{\nu}}}$ is balanced by the radiative
losses  $\mathbf{\nabla\cdot F_{\mathrm{rad}}}$.  This leads to the
height-integrated equation
\begin{equation}\label{14}
  \frac{4\sigma T_{\mathrm{c}}^{4}}{3\tau} =  \frac{9}{8}\nu\Sigma\frac{GM}{R^{3}},
 \end{equation}
where  $T_{\rm c}$ is the temperature at the midplane of the disc, $\sigma$
the Stefan-Boltzmann constant, and the optical depth of the disc is given
by:
\begin{equation}\label{15}
    \tau = \int_{0}^{H}\kappa\rho dz = \rho H\kappa = \frac{1}{2}\Sigma\kappa,
 \end{equation}
where $\kappa$ is the sum of
the electron scattering opacity
\begin{equation}\label{16}
    \kappa_{\rm es}= 0.04 \,\,\mathrm{m^{2}\,kg^{-1}},
\end{equation}
and the free-free opacity, which is given by Kramer's law
\begin{equation}\label{17}
    \kappa_{\rm ff}= \kappa_{0}\rho T_{\rm c}^{-7/2} \mathrm{m^{2}\,kg^{-1}}
\end{equation}
with
\begin{equation}\label{18}
\kappa_{0}=5\times 10^{20} \,\,\mathrm{m^{5}\,kg^{-2}\,K{^{7/2}}}.
\end{equation}
Combining Eqs. (14) and (15), we get
\begin{equation}\label{19}
    T_{\rm c}^{4} = \frac{27}{32\sigma}\rho H(\kappa_{R} + \kappa_{es})\nu\Sigma\frac{GM}{R^{3}}
\end{equation}

\subsection{The vertical structure of an accretion disc}

 The total pressure is the sum of gas and radiation pressure
\begin{equation}\label{20}
   P (\rho,T_{\rm c}) = \frac{\rho k_{\mathrm{B}}T_{\mathrm{c}}}{m_{\mathrm{p}}\bar{\mu}} + \frac{4\sigma T_{\rm c}^{4}}{3c}
\end{equation}
where  $k_{\mathrm{B}}$ is Boltzmann's constant, $\bar{\mu} = 0.62$ the mean
molecular weight for a fully ionised gas, $m_{\mathrm{p}}$ the mass of a proton,
and $c$ the speed of light,
but the pressure can also be expressed using Eq. (\ref{hydro-eq}) for hydrostatic
equilibrium.  We can thus express the scale height in terms of the total
pressure
  \begin{equation}\label{21}
  H =  \left( \frac{ k_{\rm B}T_\mathrm{c}R^{3}}{m_{\mathrm{p}}\bar{\mu}GM} + \frac{4\sigma T_{\rm c}^{4}R^{3}}{3c\rho  GM}\right)^{1/2}.
\end{equation}
The viscous stress tensor (Paper 1, Eq. 30) gives us
\begin{equation}\label{22}
  f_{R\phi} = \frac{3}{4} \frac{\Sigma \nu}{H} \left(\frac{GM}{R^3}\right)^{1/2}
= \alpha_{\rm ss} P,
\end{equation}
but pressure and density are related according to Eq. (\ref{hydro-eq}), so we
can solve for the density of the gas
\begin{equation}\label{23}
   \rho = \frac{3}{4\alpha_{\mathrm{ss}}}\frac{\nu\Sigma}{H^{3}}
\left(\frac{GM}{R^{3}}\right)^{-1/2}.
\label{density}
\end{equation}

Combining Eqs. (\ref{hydro-eq}), (\ref{ang-mom1}) and (\ref{density}) and using
$y = \nu \Sigma$  we get
  \begin{eqnarray}\label{diff_eq}
     \frac{dy}{dR} = \frac{\dot{M}}{6\pi R} - \frac{y}{2 R} - \epsilon\left[\frac{4\mu^{2}\gamma_{dyn}y}{3\mu_{0}H R^{3/2}}\right]^{1/2}(GM)^{-1/4}R^{-3/2} \nonumber\\ - \frac{4\mu^{2}\gamma}{3\mu_{0}\sqrt{GM}} R^{-9/2}\left[1 - \left(\frac{R}{R_{\rm c}}\right)^{3/2}\right],
\end{eqnarray}
where
 \begin{equation}\label{25}
   R_{\rm c}= \left(\frac{GMP_{\rm spin}^{2}}{4\pi^{2}}\right)^{1/3}\simeq 1.5\times 10^{6}
P_\mathrm{spin}^{2/3}M_{1}^{1/3}\,\mbox{m},
\end{equation}
is the co-rotation radius, at which the Keplerian  angular velocity is the same
as the stellar angular velocity.  Here
$P_{\rm spin} = \frac{2\pi}{\Omega_{s}}$  is the spin period of the star
and $M_{1} = \frac{M}{M_{\odot}}$.
As $R \to \infty$
\begin{displaymath}
y \to  \frac{\dot{M}}{3\pi},
\end{displaymath}
which we use as the outer boundary condition.

We now introduce
the dimensionless variable $\Lambda$  through
  \begin{equation}\label{26}
    y = \Lambda\dot{M},
 \end{equation}
and a dimensionless radial coordinate through
 \begin{equation}\label{27}
    R = rR_\mathrm{{A}},
 \end{equation}
where $ R_\mathrm{{A}}$ is the Alfv\'en radius, which is given by setting the magnetic pressure equal to the ram pressure
(e.g. \citealt{ss})
\begin{equation}\label{28}
    R_{\rm A} =\left(\frac{2\pi^{2}\mu^{4}}{GM\dot{M}^{2}\mu_{0}^{2}}\right)^{1/7}
\simeq 1.4\times 10^{4} {\dot{M}_{14}}^{-2/7}M_{1}^{-1/7}\mu_{16}^{4/7}\mbox{m},
\end{equation}
where $\dot{M}_{\mathrm{14}}$ represents the mass transfer rate in units of
$10^{14}\,\mathrm{kg\,s^{-1}}$, and $\mu_{\mathrm{16}}$ is the stellar
magnetic moment in units of $10^{16}$\,T\,m$^3$.
The Alfv\'en radius should not be taken as the location of the inner edge
of our discs, since this is a quantity that comes out of our solutions, but
merely as a convenient length unit.
Furthermore we introduce the fastness parameter
\begin{equation}\label{29}
\omega_{\rm s}= \left(\frac{R_{A}}{R_{\rm c}}\right)^{3/2} =
0.36M_{\mathrm{1}}^{-5/7}\dot{M}_{\mathrm{14}}^{-3/7}\mu_{\mathrm{16}}^{6/7}
\left(\frac{P_{\rm spin}}{4.8\,\mbox{ms}}\right)^{\mathrm{-1}}.
\end{equation}
We can now write  Eq.  (\ref{diff_eq}) as:
\begin{eqnarray}\label{diff_eq_2}
     \frac{d\Lambda}{dr} = \frac{1}{6\pi r} - \frac{\Lambda}{2 r} -
\epsilon\left[\frac{4\mu^{2}\gamma_{dyn}\Lambda}{3\mu_{0}H \dot{M}}\right]^{1/2}
(GM)^{-1/4}R_{A}^{-5/4}r^{-9/4} \nonumber\\ -
\frac{4\mu^{2}\gamma}{3\mu_{0}\sqrt{GM}\dot{M}} R_{A}^{-7/2}r^{-9/2}
\left[1 - \omega_{\rm s}r^{3/2}\right],
\end{eqnarray}
\section{Regional disc structure equations}
In order to  solve Eq. (\ref{diff_eq_2}) numerically we need to determine $H$ for the
different regions in the disc.
With gas pressure and Kramer's opacity
we get
\begin{equation}\label{31}
    H(r) = \left(\frac{243\kappa_{0}}{512\sigma}\right)^{1/20}
\alpha_{\rm ss}^{-1/10}
    \left(\frac{k_{\rm B}(rR_{\rm A})^{3}}{m_{p}\bar{\mu}GM}\right)^{3/8}
\left(\dot{M}\Lambda\right)^{3/20},
\end{equation}
and for the gas pressure and electron scattering opacity
\begin{equation}\label{32}
    H(r) = \left(\frac{81\kappa_{\rm es}}{128\sigma\alpha_{\rm ss}}\right)^{1/10}
    \left(\frac{k_{\rm B}}{m_{p}\bar{\mu}}\right)^{2/5}
\left(\frac{(rR_{\rm A})^{3}}{GM}\right)^{7/20}
    \left(\dot{M}\Lambda\right)^{1/5}.
\end{equation}
Finally with radiation pressure and electron scattering
the scale height is
\begin{equation}\label{h_rad_elec}
 H = \frac{9\kappa_{\rm es}}{8c}\nu\Sigma = \frac{9\kappa_{\rm es}}{8c}(\dot{M}\Lambda)
\end{equation}
which is essentially independent of R. This comes from the fact that for
radiation pressure $T_{\rm c}^4 \,\sim \,\nu\Sigma^{2} M R^{-3}.$

We can now write down the self-similar structure of the different regions of
the accretion disc in terms of the function $\Lambda(r)$.
The toroidal magnetic field is given by the same expression
\begin{equation}\label{62}
    B_{\phi,\,\mathrm{shear}} = -3.6\times 10^{3}\gamma M_{1}^{3/7}\dot{M}_{14}^{6/7}\mu_{16}^{-5/7}\left(1 -\omega_{s}r^{3/2}\right)r^{-3} \mathrm{T}
\end{equation}
everywhere in the disc, but for all other quantities we have
to handle the different regions of the disc separately.

In the outer disc where $P_{\rm g}\gg P_{\rm rad},$  and
$\kappa_{\rm ff}\gg\kappa_{\rm es}$
we have that:
 \begin{equation}\label{35}
   \Sigma = 1.6\times 10^{6}\bar{\mu}^{3/4}\alpha_{\rm ss}^{-4/5}M_{1}^{5/14}\dot{M}_{14}^{32/35}\mu_{16}^{-3/7}\Lambda(r)^{7/10}r^{-3/4} \mathrm{kg\,m^{-2}}
\end{equation}
\begin{equation}\label{36}
  T_{\rm c}= 8.1\times 10^{7}\bar{\mu}^{1/4}\alpha_{\rm ss}^{-1/5}M_{1}^{5/14}\dot{M}_{14}^{18/35}\mu_{16}^{-3/7}\Lambda(r)^{3/10}r^{-3/4}\mathrm{K}
 \end{equation}
\begin{equation}\label{37}
 \frac{H}{R}= 8.1\times 10^{-3}\bar{\mu}^{-3/8}\alpha_{\rm ss}^{-1/10}M_{1}^{-11/28}\dot{M}_{14}^{4/35}\mu_{16}^{1/14}\Lambda(r)^{3/20}r^{1/8}
\end{equation}
\begin{eqnarray}\label{38}
   \rho_{\rm c} = 7.1\times 10^{3}\bar{\mu}^{9/8}\alpha_{\rm ss}^{-7/10}M_{1}^{25/28}\dot{M}_{14}^{38/35}\mu_{16}^{-15/14}\nonumber\\
   \Lambda(r)^{11/20}r^{-15/8} \mathrm{kg\,m^{3}}
\end{eqnarray}
\begin{equation}\label{39}
   \tau =  5.9\times 10^{2} \bar\mu \alpha_{\rm ss}^{-4/5}\dot{M}_{14}^{1/5}\Lambda(r)^{1/5}
\end{equation}
\begin{equation}\label{40}
    \nu = 6.4\times 10^{7}\bar{\mu}^{-3/4}\alpha_{\rm ss}^{4/5}M_{1}^{-5/14}\dot{M}_{14}^{3/35}\mu_{16}^{3/7}\Lambda(r)^{3/10}r^{3/4}\mathrm{m^{2}\,s^{-1}}
\end{equation}
\begin{eqnarray}\label{41}
    v_{R} = 7.4\times 10^{2}\bar{\mu}^{-3/4}\alpha_{\rm ss}^{4/5}M_{1}^{-3/14}\dot{M}_{14}^{13/35}\mu_{16}^{-1/7}\nonumber\\
    \Lambda(r)^{-7/10}r^{-1/4} \mathrm{m\,s^{-1}}
\end{eqnarray}
\begin{eqnarray}\label{42}
B_{\phi,\mathrm{dyn}} = 7.7\times 10^{4}\epsilon\bar{\mu}^{3/16}\gamma_{\rm dyn}^{1/2}\alpha_{\rm ss}^{1/20}M_{1}^{5/8}\dot{M}_{14}^{4/5}\mu_{16}^{-3/4}\nonumber\\
\Lambda(r)^{17/40}r^{-21/16} \mathrm{T}
\end{eqnarray}
The transition from the outer to the middle region occurs where
$\kappa_{\rm es}
 = \kappa_{\rm ff}$.\,
Assuming that $\Lambda\simeq 1/3\pi$ at this point we have that the transition
radius is
\begin{displaymath}
 r_{OM} = 47\bar{\mu}^{-1/3}M_{\mathrm{1}}^{10/21}\dot{M}_{14}^{20/21}
\mu_{16}^{-4/7},
\end{displaymath}
which is independent of $\alpha_{\rm ss}.$

In the middle disc, where  $P_{\rm g}\gg P_{\rm rad},$  and
$\kappa_{\rm es}\gg\kappa_{\rm ff}$, we have that:
 \begin{equation}\label{44}
   \Sigma = 7.1\times10^{5}\bar{\mu}^{4/5}\alpha_{\rm ss}^{-4/5}M_{\mathrm{1}}^{2/7}\dot{M}_{14}^{27/35}\mu_{16}^{-12/35}\Lambda(r)^{3/5}r^{-3/5} \mathrm{kg\,m^{-2}}
\end{equation}
\begin{equation}\label{45}
  T_{\rm c}= 1.8\times 10^{8}\bar{\mu}^{1/5}\alpha_{\rm ss}^{-1/5}M_{1}^{3/7}\dot{M}_{14}^{23/35}\mu_{16}^{-18/35}\Lambda(r)^{2/5}r^{-9/10}\mathrm{K}
  \end{equation}
\begin{equation}\label{46}
 \frac{H}{R}= 1.2\times 10^{-2} \bar{\mu}^{-2/5}\alpha_{\rm ss}^{-1/10}M_{1}^{-5/14}\dot{M}_{14}^{13/70}\mu_{16}^{1/35}\Lambda(r)^{1/5}r^{1/20}
\end{equation}
\begin{eqnarray}\label{47}
   \rho_{\rm c} = 2.0\times 10^{3}\bar{\mu}^{6/5}\alpha_{\rm ss}^{-7/10}M_{1}^{11/14}\dot{M}_{14}^{61/70}\mu_{16}^{-33/35}\nonumber\\
   \Lambda(r)^{2/5}r^{-33/20} \mathrm{kg\,m^{-3}}
\end{eqnarray}
\begin{equation}\label{48}
   \tau =  1.4\times 10^{4}\bar{\mu}^{13/10}\alpha_{\rm ss}^{-4/5}M_{1}^{2/7}\dot{M}_{14}^{27/35}\mu_{16}^{3/35}\Lambda(r)^{3/5}r^{-3/5}
\end{equation}
\begin{equation}\label{49}
    \nu = 1.5\times 10^{8}\bar{\mu}^{-4/5}\alpha_{\rm ss}^{4/5}M_{1}^{-2/7}\dot{M}_{14}^{8/35}\mu_{16}^{12/35}\Lambda(r)^{2/5}r^{3/5}\mathrm{m^{2}\,s^{-1}}
\end{equation}
\begin{equation}\label{50}
    v_{R} = 1.6\times10^{3}\bar{\mu}^{-4/5}\alpha_{\rm ss}^{4/5}M_{1}^{-1/7}\dot{M}_{14}^{18/35}\mu_{16}^{-8/35}\Lambda(r)^{-3/5}r^{-2/5} \mathrm{m\,s^{-1}}
\end{equation}
\begin{eqnarray}\label{51}
    B_{\phi,{\rm{dyn}}} = 6.3\times 10^{4}\epsilon\bar{\mu}^{1/5} \gamma_{\rm dyn}^{1/2}\alpha_{\rm ss}^{1/20}M_{1}^{17/28}\dot{M}_{14}^{107/140}\mu_{16}^{-51/70}\nonumber\\
    \Lambda(r)^{2/5}r^{-51/40} \mathrm{T}
\end{eqnarray}
The disc structure changes drastically in the inner disc where
$P_{\rm rad} \gg P_{\rm g}$ and $\kappa_{\rm es} \gg
\kappa_{\rm ff}$,
where we have that
\begin{equation}\label{54}
   \Sigma = 9.6\times 10^{1}\alpha_{\rm ss}^{-1}M_{\mathrm{1}}^{-5/7}\dot{M}_{14}^{-10/7}\mu_{16}^{6/7}\Lambda(r)^{-1}r^{3/2} \mathrm{kg\,m^{-2}}
\end{equation}
\begin{equation}\label{55}
  T_{\rm c}= 1.9\times 10^{7}\alpha_{\rm ss}^{-1/4}M_{1}^{5/28}\dot{M}_{14}^{3/28}\mu_{16}^{-3/14}r^{-3/8}\mathrm{K}
  \end{equation}
\begin{equation}\label{56}
 \frac{H}{R}= 1.1M_{1}^{1/7} \dot{M}_{14}^{9/7}\mu_{16}^{-4/7}\Lambda(r)r^{-1}
\end{equation}
\begin{equation}\label{57}
   \rho_{\rm c} = 3.2\times 10^{-3}\alpha_{\rm ss}^{-1}M_{1}^{-5/7}\dot{M}_{14}^{-17/7}\mu_{16}^{6/7}\Lambda(r)^{-2}r^{3/2} \mathrm{kg\,m^{-3}}
\end{equation}
\begin{equation}\label{58}
   \tau_{es} =  1.9\alpha_{\rm ss}^{-1}M_{1}^{-5/7}\dot{M}_{14}^{-10/7}\mu_{16}^{6/7}\Lambda(r)^{-1}r^{3/2}
\end{equation}
\begin{equation}\label{59}
    \nu = 1.0\times 10^{12}\alpha_{\rm ss}M_{1}^{5/7}\dot{M}_{14}^{17/7}\mu_{16}^{-6/7}\Lambda(r)^{2}r^{-3/2}\mathrm{m^{2}\,s^{-1}}
\end{equation}
\begin{equation}\label{60}
    v_{R} = 1.2\times10^{7}\alpha_{\rm ss}M_{1}^{6/7}\dot{M}_{14}^{19/7}\mu_{16}^{-10/7}\Lambda(r)r^{-5/2} \mathrm{m\,s^{-1}}
\end{equation}
\begin{equation}\label{61}
    B_{\phi,\mathrm{dyn}} = 6.6\times10^{3}\epsilon\gamma_{\rm dyn}^{1/2}
\alpha_{\rm ss}^{1/20} M_{1}^{5/14}\dot{M}_{14}^{3/14}\mu_{16}^{-3/7}r^{-3/4} \mathrm{T}
\end{equation}
It is of particular significance here that $\Sigma \propto \Lambda^{-1}$.

The transition radius  between the middle and the inner region is
estimated by equating  the gas and radiation pressures in the middle region
and approximating $\Lambda$ with $1/3\pi$
\begin{displaymath}
 r_{IM} = 12\bar{\mu}^{8/21}\alpha_{\rm ss}^{2/21}M_{\mathrm{1}}^{10/21}
\dot{M}_{14}^{22/21}\mu_{16}^{-4/7}
\end{displaymath}
We then see that the accretion disc right outside of the Alfv\'en
radius is dominated by radiation pressure only if
\begin{equation}
  \mu_{16} < 83 \bar\mu^{2/3} \alpha_{\rm ss}^{1/6} M_1^{5/6} \dot M_{14}^{11/6}
\end{equation}
This condition is not fulfilled for a conventional X-ray pulsar with
a magnetic dipole moment of $\sim 10^{20}$\,T\,m$^3$
\citep{ws}, though it can be fulfilled for a millisecond
X-ray pulsar.

 \section{Global solutions }

For our fiducial model we take a neutron star of $M = 1.4M_{\odot}$ with a
magnetic moment of $10^{16}\,\mathrm{T\,m^{3}}$ and a spin period of 4.8\,ms.
We set the dimensionless
parameters $\gamma$ and $\gamma_{\rm{dyn}}$ to 1 and 10, respectively,
while $\alpha_{\rm ss} = 0.01.$
$\gamma$,\,  $\gamma_{\rm{dyn}}$, and $\alpha_{\rm ss}$ appear only in
combinations with other parameters, and thus their values do not carry any
special significance, but the parameters $\epsilon$ and $\dot M$ appear on their
own in the equation, and therefore influence the solution in unique ways.
In particular $\dot M$ influences the fastness parameter and the
transition radii between the regions of the accretion disc.  In this
paper we consider three different accretion rates, $\dot M_{14} = 0.012$,
0.12 and 1.5.  We list the corresponding fastness parameters and transition
radii in Tab. \ref{params}.

\begin{table}
\caption{The Alfv\'en radius, fastness parameter $\omega_{\rm s}$ and the
transition radii in units of the Alfv\'en radius
for different accretion rates $\dot M_{14}$ and the fiducial neutron star.}
\label{params}
\begin{tabular}{lllll}
\hline \\
$\dot M_{14}$ & $R_{\rm A}$\, [m] & $\omega_{\rm s}$ & $r_{\rm OM}$ &
$r_{\rm MI}$\\
\hline \\
0.012 & $4.7\times 10^4$ & 0.98 & \\
0.12 & $2.4\times 10^4$ & 0.37 & 8.5 & --\\
0.8 & $1.4\times 10^4$ & 0.31 & 52 & 6.0\\
1.5 & $1.2\times 10^4$ & 0.12 & 95 & 12 \\
\hline \\
\end{tabular}
\end{table}

\subsection{Case I: $R_{\rm A} > R_{\rm OM}$}

\begin{figure}
\includegraphics[width = 8.4cm]{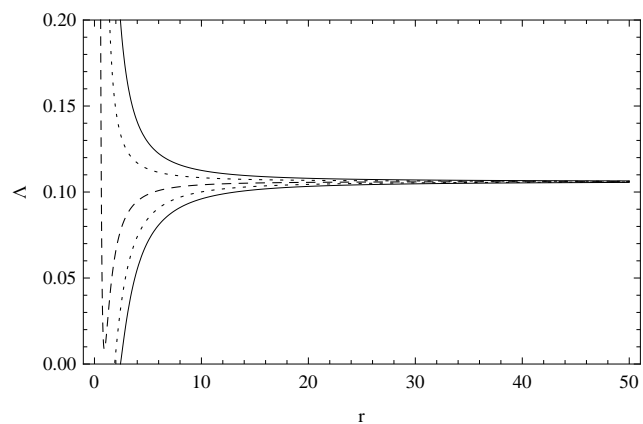}
\caption{ $\Lambda(r)$ for our fiducial neutron star with
an accretion rate of $1.2\times10^{12}$\,kg\,s$^{-1}$,
and $\epsilon= 0.1,\,0.05, \, 0,\,-0.05,\,-0.1$
from the top to the bottom.  In this case the disc
consists of only an outer region that is dominated by gas pressure and
Kramer's opacity.}
\label{Lambdafigout}
\end{figure}

\begin{figure}
\includegraphics[width = 8.4cm]{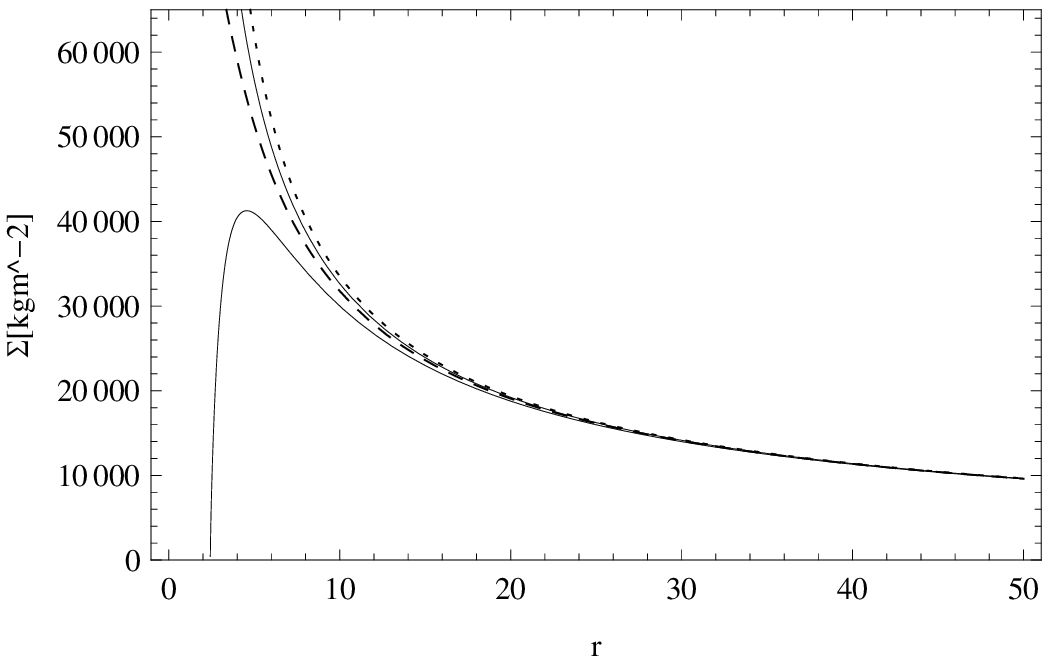}
\caption{  $\Sigma$  as a function of $r$ for our fiducial neutron star with
an accretion rate of
$1.2\times10^{12}$\,kg\,s$^{-1}$,
and with $\epsilon= 0.1,\,0.05, \, 0,\,$ and $-0.1$ from the top to the bottom.}
\label{sigmafigout}
\end{figure}

\begin{figure}
\includegraphics[width = 8.4cm]{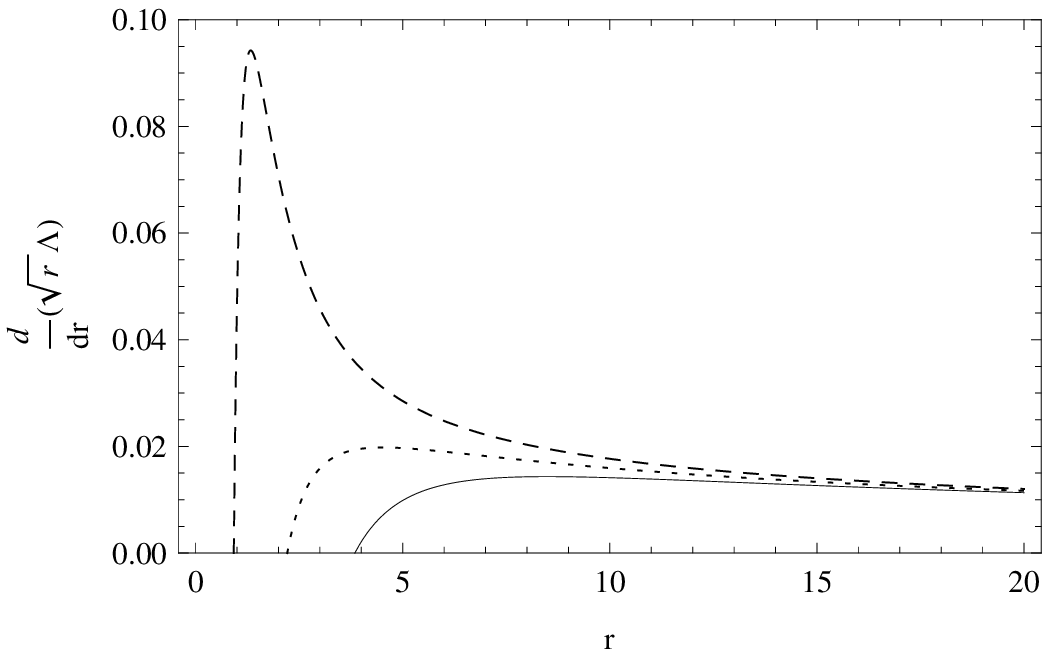}
\caption{$\frac{d}{dr}\left(\sqrt{r}\Lambda\right)$  as a function of $r$ for
the fiducial neutron star with an accretion rate of
$1.2\times10^{12}$\,kg\,s$^{-1}$,
and with $\epsilon = \, 0.1,\,\,0.05$,\, and \,$0$ from the bottom to the top.}
\label{innerfigouter}
\end{figure}

The accretion disc consists of only an outer region when
$r_{\mathrm{OM}} < 1$,
which corresponds to that $\dot M < 1.4 \times 10^{12}$\,kg\,s$^{-1}$ for our
fiducial model.  Assuming $\dot{ M} = 0.012\dot{ M}_{14}$ we get the solutions  from the
top to the bottom of
Fig. \ref{Lambdafigout} for $\epsilon = 0.1,\,0.05,\,0,\,-0.05$ and $-0.1$.
We see here that $\Lambda \to 0$ at a finite $r$ for $\epsilon = -0.1$,\,and\,$-0.05$, which we
 called case D in Paper I, since $\rho \to 0$ at this radius
(Fig. \ref{sigmafigout}).
In the other cases
$\Lambda$ grows without a bound for decreasing $r$, which we
call case V in Paper I, since then the viscous flux of angular momentum
changes sign at a finite radius where
$\frac{\mbox{d}}{\mbox{d}r}(\sqrt{r}\Lambda) = 0$ (Fig. \ref{innerfigouter}),
which we take as the inner edge of the disc.  Inside of this radius the
accretion flow is driven by the magnetic coupling to the neutron star.

\begin{figure}
\includegraphics[width = 8.4cm]{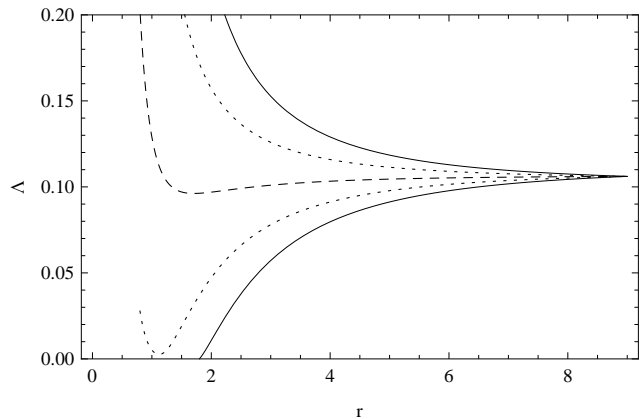}
\caption{ $\Lambda(r)$ for the middle region of an accretion disc around our
fiducial neutron star with an accretion rate of
$1.2\times 10^{13}$\,kg\,s$^{-1}$,
and with $\epsilon= 0.1,\,0.05, \, 0,\,-0.05,\,-0.1$
from the top to the bottom.}
\label{Lambdafigmiddle}
\end{figure}

\subsection{Case II: $R_{\rm IM}< R_{\rm A}<R_{\rm OM}$}

The disc has both an outer and a middle region when
$0.014 < \dot M_{14} <0.14$.
We then use the analytical solution with $\Lambda = 1/3\pi$ for the outer region
and solve Eq. (\ref{diff_eq_2}) numerically for the middle region.  For our
calculations we take $\dot M = 1.2\times 10^{13}$\,kg\,s$^{-1}$, which
yields $r_{\rm OM} \approx 8.5$.
The numerical solutions for the middle region are shown in Fig.
\ref{Lambdafigmiddle} for our fiducial neutron star with
$\epsilon = \, 0.1,\,0.05,\,0,\,-0.05$,\, and  $-0.1$.
We find a case D inner boundary for $\epsilon =-0.1$,
while the other solutions have case V
inner boundaries.
The transition from case D to case V at the $\epsilon$, that gives
the smallest possible radius for the inner edge of the disc, which varies weakly
with the accretion rate $\dot M$.

\begin{figure}
\includegraphics[width = 8.4cm]{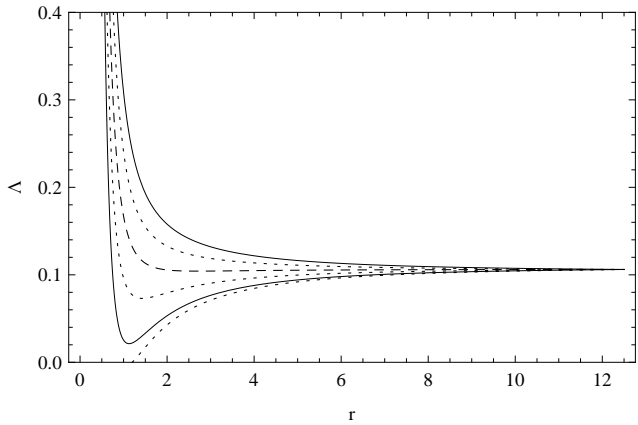}
\caption{ $\Lambda(r)$ for our fiducial neutron star with
an accretion rate of $1.5\times 10^{14}$\,kg\,s$^{-1}$,
and with $\epsilon= 0.1,\,0.05, \, 0,\,-0.05,\,-0.1,\,-0.12$ from the top to the
bottom.}
\label{Lambdafiginner}
\end{figure}

\begin{figure}
\includegraphics[width = 8.4cm]{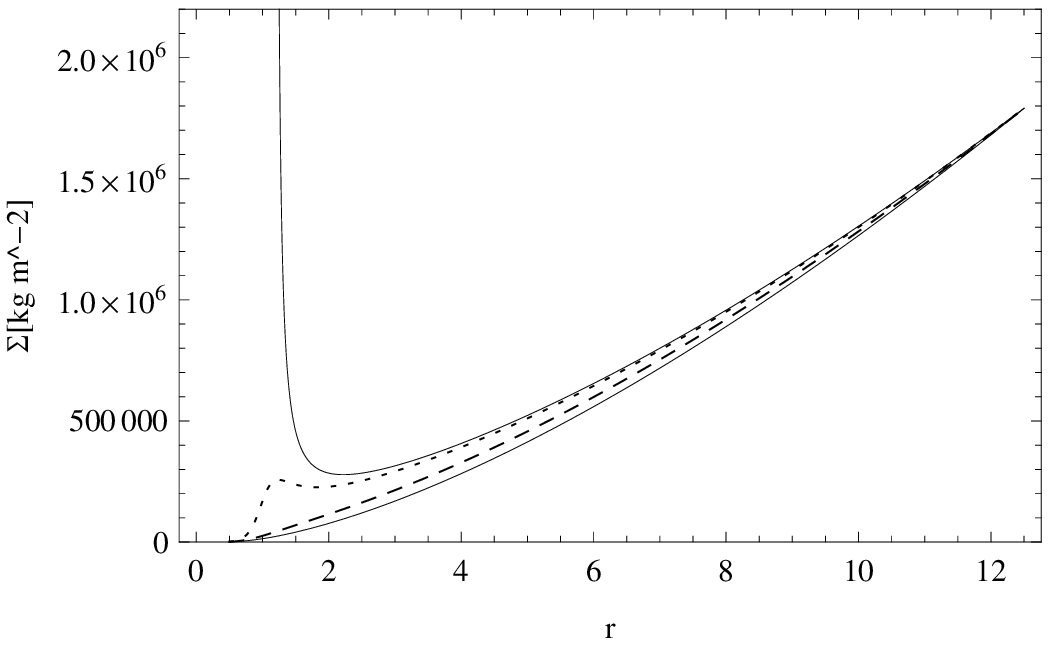}
\caption{$\Sigma$ as a function of $r$ for the fiducial  neutron star with
an accretion rate of $1.5\times 10^{14}$\,kg\,s$^{-1}$,\,
and with $\epsilon = \,0.1,\,0,\,-0.1,\,$ and $-0.12$ from the bottom
to the top.}
\label{sigmainn}
\end{figure}

\begin{figure}
\includegraphics[width = 8.4cm]{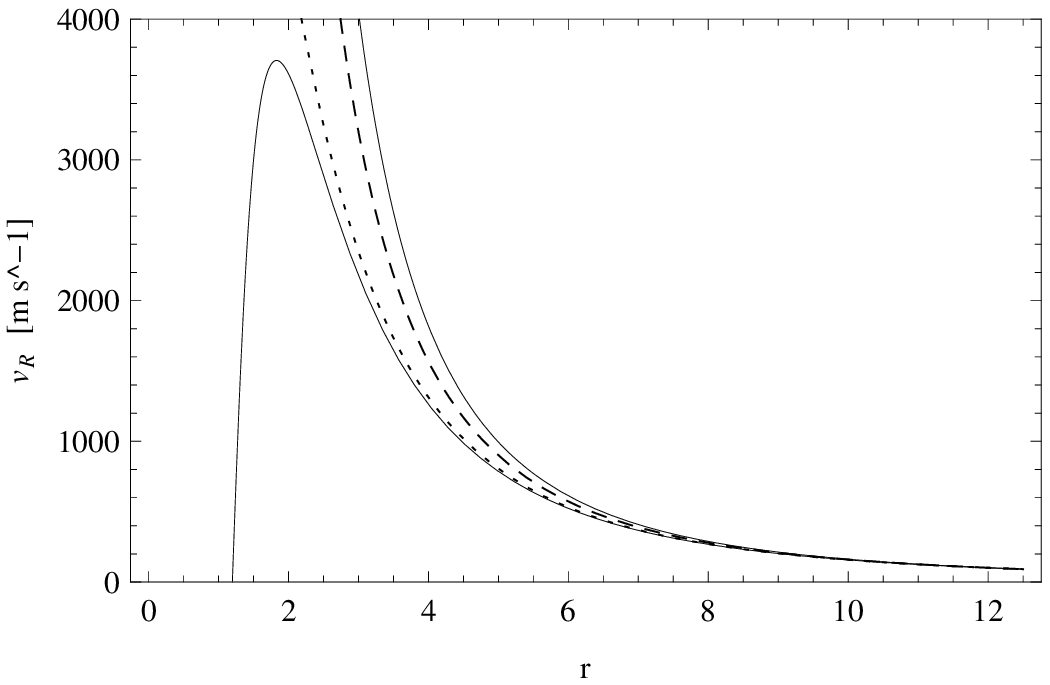}
\caption{ $V_{R}$  as a function of $r$ for the fiducial neutron star with
an accretion rate of $1.5\times 10^{14}$\,kg\,s$^{-1}$
and $\epsilon=\,0.1,\,0,\,-0.1\,$ and $-0.12$ from the top to the bottom}
\label{radialvelocityfigInn}
\end{figure}

\begin{figure}
\includegraphics[width = 8.4cm]{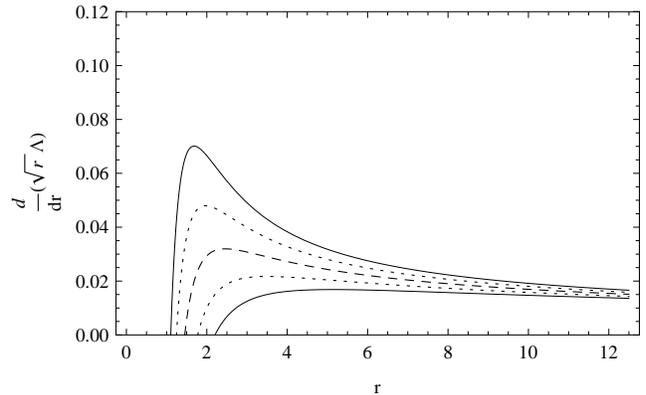}
\caption{$\frac{d}{dr}\left(\sqrt{r}\Lambda\right)$  as a function of $r$ for
the fiducial neutron star with an accretion rate of
$1.5\times 10^{14}$\,kg\,s$^{-1}$,\,
and with $\epsilon = \, 0.1,\,0.05, \,0,\,-0.05$, and $\,-0.1\,$  from the bottom to the top.}
\label{innerfigin}
\end{figure}

\subsection{Case III: $R_{\rm A} < R_{\rm IM}$}

The innermost region of the disc is dominated by radiation pressure and
electron scattering when $\dot{M}_{14} > 0.14$.
We use two different accretion rates that fulfill this condition,
$\dot{M} = 8\times 10^{13}$\,kg\,s$^{-1}$ and\,
$\dot{M} = 1.5\times 10^{14}$\,kg\,s$^{-1}$ together with our fiducial
neutron star in our calculations.
We assume that
$\Lambda = 1/3\pi$ in the outer and
middle regions of the disc, and solve Eq. (\ref{diff_eq_2}) for the inner
region starting from $r_{\rm IM} \approx 12.$
Our solutions for the inner disc region at the higher accretion rate and
with $\epsilon= 0.1,\,0.05, \, 0,\,-0.05,\,-0.1$,\, and $-0.12$ are shown in Fig.
\ref{Lambdafiginner}.
All these solutions have case V inner boundaries except the
$\epsilon = -0.12$ model.
 However because  $\Sigma \varpropto \Lambda^{-1}$  in this case we see that the density goes to infinity at the inner edge of the
 $\epsilon= -0.12 $ disc (Fig. \ref{sigmainn}), at which point the flow
stagnates (Fig. \ref{radialvelocityfigInn}). In the case V solutions, on the
other hand, we see that $\Sigma\rightarrow 0$ at a point which is inside the
inner edge of the disc (Fig. \ref{innerfigin}).

\section{Discussion}

\subsection{The inner edge of the accretion disc}

In order for the neutron star to appear as an X-ray pulsar the accretion disc
must be truncated
by the stellar magnetic field (Tab. \ref{table1})
at a radius $R_0$ outside of the neutron star.
The inner edge of the disc
can be either
of type D ($\Lambda = 0$) or type V ($\frac{\mbox{d}}{\mbox{d}r} (\sqrt{r}
\Lambda) = 0$) (see Paper I for an extensive discussion of these boundaries).
In models without a dynamo (for instance \citealt{gb}) $R_0$ is comparable to
or smaller than the Alfv\'en radius, $R_{\rm A}$, but we find that the dynamo
can make it significantly larger than $R_{\rm A}$.  This opens up the
possibility, as suggested by the referee, that a system can temporarily turn
into an X-ray pulsar during the periods when the magnetic field
generated by the dynamo is strong enough to make $R_0$ larger than the
radius of the neutron star.  SAX J1748.9-2021 is such an intermittent
millisecond X-ray pulsar \citep{alt}, which is on for hundreds of seconds at
a time.

In case $R_0$ is located in the outer or middle region of our
disc solution we find that a type D boundary coincides with the point at
which $\rho \to 0$ and $v_R \to \infty$ though the accretion rate is finite.
However a type D boundary in the inner part of the accretion disc behaves in
the opposite way; $\rho \to \infty$ and $v_R \to 0$.  Neither one of these
behaviours is entirely physical and one must keep in mind that we have
ignored the importance of the radial pressure gradient and the
inertial term in Eq. (\ref{3}) (see for instance \citealt{abr88,pop92}), and
we have not attempted to model how the plasma is entering the stellar
magnetic field lines to form a funnel flow onto the magnetic poles of the
neutron star.
One further problem with the inner region of the disc solution is that it
is known to be viscously and thermally unstable \citep{le}.  This instability
persists if the accretion disc is threaded by a stellar magnetic field
\citep{b},
but so far it has not been investigated what happens if the accretion disc
is sustaining its own magnetic field through a dynamo.

\subsection{Accretion torque}

To understand the nature of the exchange of angular momentum between the accretion
disc and the neutron star we multiply Eq. (\ref{ang-mom1}) by $2\pi R$ and integrate
it from $R_0$, the inner radius of the disc, to a large radius $R_1$.
\begin{eqnarray}
  -\dot M \left(\sqrt{GMR_1}-\sqrt{GMR_0}\right)= \nonumber\\ \int_{R_0}^{R_1} 4\pi
\frac{B_z B_{\phi,{\rm dyn}}}{\mu_0} R^2 \mbox{d}R +
\int_{R_0}^{R_1} 4\pi \frac{B_z B_{\phi,{\rm shear}}}{\mu_0} R^2 \mbox{d}R\nonumber\\
-3\pi \left(\nu \Sigma\right)_{R_1} \sqrt{GMR_1}
+ 3\pi \left(\nu \Sigma\right)_{R_0} \sqrt{GMR_0}.
\label{torque}
\end{eqnarray}
The left-hand-side is the difference between the angular momentum that is
advected through the inner edge of the accretion disc and that which is fed into
the disc at its outer edge and the right-hand-side describes the contribution of  magnetic and viscous torques
to the angular momentum balance.  Notice that the term $3\pi(\nu \Sigma)_{R_1}
\sqrt{GMR_1}$ describes the viscous tension at the outer edge of the
disc, which will not be considered further in this paper.
The material torque of the inner edge of the disc on the neutron star is
\begin{equation}\label{58}
    N_{adv} = \dot{M}\sqrt{GMR_{0}} = 1.4\times10^{26}\mu_{16}^{2/7}M_{1}^{3/7}\dot{M}_{14}^{6/7}r_{0}^{1/2}.
    \end{equation}
It increases as the accretion rate increases (Tab.\ref{table1}).
Then we have the magnetic torques on the neutron star, which we have divided
into
one part due to the shear
  \begin{eqnarray}\label{59}
    N_{\rm shr} = - \int_{R_0}^{R_1} 4\pi
\frac{B_z B_{\phi,{\rm shear}}}{\mu_0} R^2 \mbox{d}R \approx \nonumber \\
3.6\times 10^{26}
\gamma \mu_{16}^{2/7}M_{1}^{3/7}\dot{M}_{14}^{6/7}\nonumber\\
    \int_{r_{0}}^{\infty}\left[r^{-4}\left(1 - \omega_{s}r^{3/2}\right)\right]
\mbox{d}r
  \end{eqnarray}
and a second part due to the dynamo
\begin{equation}\label{60}
    N_{\rm dyn}= - \int_{R_0}^{R_1} 4\pi
\frac{B_z B_{\phi,{\rm dyn}}}{\mu_0} R^2 \mbox{d}R.
\end{equation}
We split up the latter integral into one integral for each of
the three different regions of the disc
\begin{eqnarray}
    N_{\rm dyn,outer} = 7.7\times 10^{27}\epsilon\gamma_{dyn}^{1/2}\alpha_\mathrm{ss}^{1/20}\bar{\mu}^{3/16}\mu_{16}^{1/4}M_{1}^{5/8}\dot{M}_{14}^{4/5}\nonumber\\
    \int_{\rm outer}\Lambda^{17/40}r^{-37/16}dr,
\end{eqnarray}
\begin{eqnarray}
     N_{\rm dyn,mid} = 6.3\times 10^{27}\epsilon\gamma_{dyn}^{1/2}\alpha_\mathrm{ss}^{1/20}\bar{\mu}^{1/5}\mu_{16}^{19/70}M_{1}^{17/28}\dot{M}_{14}^{107/140}\nonumber\\
    \int_{\rm middle}\Lambda^{2/5}r^{-91/40}dr,
\end{eqnarray}
and
\begin{eqnarray}
     N_{\rm dyn,inner} = 6.6\times 10^{26}\epsilon\gamma_{dyn}^{1/2}\mu_{16}^{4/7}M_{1}^{5/14}\dot{M}_{14}^{3/14}
    \int_{\rm inner}r^{-7/4}dr.
\end{eqnarray}
Finally we have the boundary terms describing the viscous stresses on the
inner and the outer edges of the accretion disc.
The viscous stress at the outer edge of the disc is responsible for
removing the angular momentum that is transported outwards through the
accretion disc, and does not affect the angular momentum evolution of the
neutron star.  The viscous stress at the inner edge of the disc on the other
hand can result in a torque that transports
angular momentum away from the neutron star
\begin{eqnarray}\label{61}
    N_{\rm vis} = - 3\pi \left(\nu \Sigma\right)_{R_0} \sqrt{GMR_0} =\nonumber\\
-1.29\times10^{27}\mu_{16}^{2/7}M_{1}^{3/7}\dot{M}_{14}^{6/7}\Lambda(r_{0}) r_{0}^{1/2}
\end{eqnarray}
This term is usually neglected in accretion disc theory since the standard
accretion disc solution has a case D inner boundary. But it
can transport angular momentum from the neutron star to the disc when we have a
case V inner boundary.

We report on all the components and the total sum of the torques on our fiducial neutron star
in Tab.\ref{table1}.
The main factor determining the total torque is the accretion rate.
In Paper I we found that the dominant contributor to the torque is the
internal disc dynamo.  This torque is still strong, but now $N_{\rm adv}$
and $N_{\rm vis}$ are comparable in strength, though these make opposing
contributions to the net torque, while the contribution from
$N_{\rm sh}$ is still small.  There is therefore no simple dependence of
$N_{\rm total}$ on $\epsilon$, though the dynamo is indirectly affecting
$N_{\rm adv}$
and $N_{\rm vis}$ by causing the inner disc edge to move outwards compared
to the \citet{gb} model.  $N_{\rm vis}$ is also highly sensitive to the
nature of the inner disc boundary, and in particular it vanishes for a D
inner boundary, which occurs at a sufficiently negative $\epsilon$, in
which case the net torque is determined by the balance between $N_{\rm dyn}$
and $N_{\rm adv}$.

 \begin{table*}
 \caption{The inner\, edge of the accretion disc and its torque on the
fiducial neutron star for different accretion rate.}
\label{table1}
\begin{footnotesize}
\begin{tabular}{llllllllll}
  \hline
  $\dot M_{14}$ & $\epsilon$ &$R_{0}\,[m]$& $N_{\rm dyn,outer}$\,[Nm] &
$N_{\rm dyn,Middle}$\,[Nm] & $N_{\rm dyn,inner}$\,[Nm] &$N_{\rm sh}$\,[Nm]  &
$N_{\rm adv}$\,[Nm] &    $N_{\rm vis}$\,[Nm] & $N_{\rm total}$\,[Nm]\\
   \hline
   $0.012$ & $0.1$ & $1.8\times10^{5}$ & $3.4\times10^{24}$ & -- & -- & $-7.8\times10^{23}$& $7.0\times10^{24}$ &  $-7.4\times10^{24}$ & $2.2\times10^{24}$ \\
   & $0.05$ & $9.9\times10^{4}$ & $3.7\times10^{24}$ & -- & -- & $-1.7\times10^{24}$  & $5.2\times10^{24}$ &  $-6.7\times10^{24}$ & $5.0\times10^{23}$\\
   & $0$ & $4.7\times10^{4}$ & $0$ & -- & -- & $-3.1\times10^{24}$  & $3.6\times10^{24}$ &  $-3.4\times10^{22}$  &  $4.7\times10^{23}$ \\
   & $-0.05$ & $8.8\times10^{4}$ & $-2.9\times10^{24}$ & -- & -- & $-1.9\times10^{24}$ &  $4.9\times10^{24}$ &  $0$ & $1.0\times10^{23}$\\
   & $-0.1$ & $1.2\times10^{5}$ & $-4.1\times10^{24}$ & -- & -- & $-1.4\times10^{24}$& $5.6\times10^{24}$ & $0$  &  $1.0\times10^{23}$\\
  \hline
  $0.12$ & $0.1$ & $9.4\times10^{4}$ & $8.4\times10^{25}$ & $1.4\times10^{25}$ &-& $-2.1\times10^{24}$  &  $5.1\times10^{25}$ &  $-1.0\times10^{26}$ & $4.7\times10^{25}$ \\
   &  $0.05$  & $6.0\times10^{4}$ & $4.1\times10^{25}$ & $1.5\times10^{25}$ &-& $-3.2\times10^{24}$  &  $4.1\times10^{25}$ & $-7.7\times10^{25}$ & $1.7\times10^{25}$ \\
   &  $0$ & $3.0\times10^{4}$ & $0$ & $0$ & - & $-7.4\times10^{23}$ &  $2.9\times10^{25}$ & $-2.4\times10^{25}$& $4.3\times10^{24}$\\
   &  $-0.05$  & $3.2\times10^{4}$ & $-4.0\times10^{25}$& $-2.6\times10^{25}$ & - & $-2.0\times10^{24}$ &  $2.9\times10^{25}$ &  $-6.7\times10^{22}$ & $-3.9\times10^{25}$ \\
   &  $-0.1$  & $4.8\times10^{4}$ & $-7.9\times10^{25}$ & $-3.0\times10^{25}$ & - & $-3.7\times10^{24}$ &  $3.6\times10^{25}$ &  $0$ &$-7.7\times10^{25}$ \\
   \hline
   \hline

    $0.8$ &$0.1$  & $3.2\times10^{4}$ & $4.9\times10^{24}$ & $3.1\times10^{23}$ & $8.2\times10^{25}$ & $-1.0\times10^{25}$  & $2.0\times10^{26}$ &  $-2.7\times10^{26}$ & $7.2\times10^{24}$\\
   & $0.05$ & $1.6\times10^{4}$ & $2.5\times10^{24}$ & $3.2\times10^{25}$ & $1.1\times10^{26}$ & $2.7\times10^{25}$ & $1.4\times10^{26}$ & $-2.5\times10^{26}$ & $6.2\times10^{25}$ \\
   & $0$ & $1.7\times10^{4}$ & $0$ & $0$ & $0$ & $1.4\times10^{25}$ &  $1.5\times10^{26}$ &  $-1.6\times10^{26}$ & $4.0\times10^{24}$ \\
   & $-0.05$ & $2.0\times10^{4}$ & $-2.4\times10^{24}$ & $8.3\times10^{22}$ & $-7.7\times10^{25}$ & $2.5\times10^{23}$ &$1.6\times10^{26}$ & $-6.2\times10^{25}$ & $1.9\times10^{25}$ \\
   & $-0.1$  & $1.7\times10^{4}$ & $-4.8\times10^{24}$ & $-1.3\times10^{25}$ & $-1.8\times10^{26}$ & $1.4\times10^{25}$  & $1.4\times10^{26}$ &  $0$ & $-4.4\times10^{25}$ \\
   & $-0.12$  & $1.8\times10^{4}$ & $-5.8\times10^{24}$ & $-8.4\times10^{24}$ & $-2.0\times10^{26}$ & $-3.2\times10^{24}$   & $1.5\times10^{26}$ &  $0$& $-6.7\times10^{25}$\\
  $1.5$ &$0.1$  & $2.6\times10^{4}$ & $2.1\times10^{24}$ & $2.5\times10^{25}$ & $1.1\times10^{26}$ & $-5.6\times10^{24}$  & $3.3\times10^{26}$ &  $-4.5\times10^{26}$ & $1.2\times10^{25}$\\
   & $0.05$ & $2.1\times10^{4}$ & $1.0\times10^{24}$ & $1.3\times10^{25}$ & $7.1\times10^{25}$ & $4.6\times10^{24}$ & $2.9\times10^{26}$ & $-4.2\times10^{26}$ & $-4.0\times10^{25}$ \\
   & $0$ & $1.8\times10^{4}$ & $0$ & $0$ & $0$ & $1.7\times10^{25}$ &  $2.7\times10^{26}$ &  $-3.2\times10^{26}$ & $-3.3\times10^{25}$ \\
   & $-0.05$ & $1.6\times10^{4}$ & $-1.1\times10^{24}$ & $-1.3\times10^{25}$ & $-9.1\times10^{25}$ & $3.7\times10^{25}$ &$2.6\times10^{26}$ & $-3.2\times10^{26}$ & $-1.3\times10^{26}$ \\
   & $-0.1$  & $1.5\times10^{4}$ & $-2.1\times10^{24}$ & $-2.5\times10^{25}$ & $-2.0\times10^{26}$ & $5.5\times10^{25}$  & $2.5\times10^{26}$ &  $-1.1\times10^{25}$ & $6.7\times10^{25}$ \\
   & $-0.12$  & $1.3\times10^{4}$ & $-2.5\times10^{24}$ & $-3.0\times10^{25}$ & $-2.4\times10^{26}$ & $5.8\times10^{25}$   & $2.5\times10^{26}$ &  $0$& $3.6\times10^{25}$\\
  \hline
  \hline
\end{tabular}
\end{footnotesize}
\end{table*}

\subsection{Observed properties of millisecond X-ray Pulsars}

The largest spin up rate recorded so far for a millisecond X-ray pulsar is
$\sim 10^{-12}$\,Hz\,s$^{-1}$ during the December 2004 outburst of
IGR J00291+5934 \citep{bur07}, but one should keep in mind that
there are large uncertainties in the spin variations that have
been reported for the millisecond X-ray pulsars.  For instance
\citet{bur06} reported
$\dot \nu$s between $-7.6\times 10^{-14}$ and $4.4\times 10^{-13}$\,Hz\,s$^{-1}$
for SAX J1808.4-3658, but \citet{hm} noted that the measurements of this source
are plagued by large variations in the pulse shape, and put an upper limit
of $2.5\times 10^{-14}$\,Hz\,s$^{-1}$ on the spin variations at any instant.
Some of the timing noise in the millisecond X-ray pulsars can be explained by
the motion of the hot spot on
the neutron star \citep{pat}, and may thus not correspond to real changes in
the spin period of the neutron star.

The spin variation corresponds to a torque
 \begin{equation}
 N = 2\pi\dot{\nu}I = 6.3\times10^{25}\dot{\nu}_{13}I_{38} \,\mathrm{Nm},
\end{equation}
where $I_{38}$ is the moment of inertia of the neutron star measured
in $10^{38}$\,kg\,m$^{2}$, and  $\dot \nu_{13}$ is the spin derivative in
units of $10^{-13}$\,Hz\,s$^{-1}$.
In the \citet{gb} model, the spin variation of IGR J00291+5934 would imply
then an
accretion rate of at least $\sim 10^{14}$\,kg\,s$^{-1}$, which is uncomfortably
large, but our
Tab. \ref{table1} shows that it is possible to enhance the torque severalfold
by coupling the stellar magnetic field to the magnetic field generated by an
internal dynamo in the disc, and through the fact that the inner edge of the
accretion disc is then located outside the Alfv\'en radius, which
increases the amount of angular momentum that is advected onto the neutron
star.  Thus it is
possible that a lower accretion rate can produce the necessary torque, though
we have not attempted to fit our model to this system.

\subsection{The validity of the model}

A significant part of the luminosity is not generated in the accretion disc
outside of $R_0$, but rather it is released as the accreting matter falls onto
the neutron star.  We estimate that this luminosity is
\begin{equation}
  L_{\rm NS} = GM\dot M\left(\frac{1}{R_{\rm NS}} - \frac{1}{R_0}\right) +
\frac{1}{2}\frac{GM\dot M}{R_0} = GM\dot M\left(\frac{1}{R_{\rm NS}} -
\frac{1}{2R_0}\right),
\end{equation}
where we have assumed that both the potential energy difference between the
inner edge of the accretion disc and the surface of the neutron star
as well as the
kinetic energy associated with the Keplerian motion at the inner edge of
the accretion disc is released at the surface of the neutron star.
 Consequently $GM\dot M/2R_{\rm NS} \le L_{\rm NS} \le GM \dot M/R_{\rm NS}$.

The neutron star can then contribute to the local heating in the accretion
disc through its irradiation of the disc.
Assuming that the neutron star is a point source and the accretion disc is
flaring, i.e. $\mbox{d}\ln H/\mbox{d} \ln R > 0$, the irradiation flux on
the accretion disc is  at most \citep{fkr}
\begin{eqnarray}
  F_{\rm irr} = \frac{L_{\rm NS}}{4\pi R^2} \frac{H}{R}
\left(\frac{\mbox{d}\ln H}{\mbox{d}\ln R} - 1\right)
\left(1-\beta\right) = \nonumber \\
\frac{GM\dot M}{4\pi R_{\rm NS} R^2}\frac{H}{R}
\left(\frac{\mbox{d}\ln H}{\mbox{d}\ln R} - 1\right)
\left(1-\beta\right),
\end{eqnarray}
where $\beta$ is the disc albedo (for a more complete treatment
taking into account shadowing by the disc see \citealt{ada86}).
The local flux generated by the viscous heating in the accretion disc is on
the other hand
\begin{equation}
  F_{\rm accr} = \frac{9}{8} \frac{GM\nu \Sigma}{R^3}.
\end{equation}
We thus find that
\begin{equation}
  \frac{F_{\rm irr}}{F_{\rm accr}} = \frac{2}{3} \frac{H}{R_{\rm NS}}
\left(\frac{\mbox{d}\ln H}{\mbox{d}\ln R} -1\right) \left(1-\beta\right),
\end{equation}
and since $\mbox{d}\ln H/\mbox{d}\ln R - 1\sim 0.1$ we find that irradiation
dominates for $H \ga 15 R_{\rm NS}$ if $\beta = 0$.
 Consequently irradiation dominates over viscous heating for $R \ga 10^{4}$\,km.
Thus the outer part of the accretion disc is hotter than is predicted by our
model, which can be observed as an increased UV flux \citep{vrtilek}.  Thus
we have underestimated the pressure in the outer disc, and since
we have used the \citet{bs} prescription $\alpha_{\rm ss} P$
to estimate $B_{\phi, {\rm dyn}}$,
we have also underestimated $B_{\phi, {\rm dyn}}$ and the torque from the
outer disc on the neutron star.  However, this torque is small in comparison
to the total torque on the neutron star since $B_z B_{\phi, {\rm dyn}}$ is a
rapidly decreasing function of $R$, and thus the effect of the irradiation
of the outer disc does not significantly change our results.
Irradiation is unlikely to significantly affect the inner disc, but, as
pointed out above, other effects, such as radial advection and the transition
from a disc flow to a funnel flow, that are not considered in this paper
become important in the inner disc.  We will investigate these effects in the
future.

 \section{Conclusions}

The accretion rates that have been observed in millisecond X-ray pulsars cover
several orders of magnitude, and the highest accretion rates are sufficient
that the innermost part of the accretion disc is dominated by radiation
pressure and electron scattering.  For these reasons we have followed the
approach by \citet{bs} and divided our
disc model in three regions, an outer region dominated by gas pressures and
Kramer's opacity, a middle region dominated by gas pressure and electron
scattering, and an inner region dominated by radiation pressure and electron
scattering.

The large spin changes that have been observed in
some of the accreting millisecond X-ray pulsars are
difficult to explain in the model by \citet{gb}, but the inclusion of a disc
dynamo
produces a significant enhancement of the torque between the accretion disc
and the neutron star.  This increase is not only due to the coupling
between the magnetic fields of the neutron star and the accretion disc as such,
but
also to the fact that the
accretion disc is truncated at a larger radius
thus increasing the lever arm between the edge of the disc and the neutron
star.

\section*{Acknowledgments}
SBT thanks the Department of Physics at the University of Gothenburg for their
hospitality and support during a part of this project.
 SBT is supported in part by the Swedish Institute (SI) under their Guest Scholarship Program.
This research has made use of NASA's Astrophysics Data System. We
thank an anonymous referee for comments that have improved the quality of the
paper.  We have also benefitted from discussions with Dr. Jonathan Ferreira.

\end{document}